# Security Services Using Blockchains: A State of the Art Survey[1]

Tara Salman, Student Member, IEEE, Maede Zolanvari, Student Member, IEEE, Aiman Erbad, Member, IEEE, Raj Jain, Fellow, IEEE, and Mohammed Samaka, Member, IEEE

*Abstract*—This article surveys blockchain-based approaches for several security services. These services include authentication, confidentiality, privacy and access control list (ACL), data and resource provenance, and integrity assurance. All these services are critical for the current distributed applications, especially due to the large amount of data being processed over the networks and the use of cloud computing. Authentication ensures that the user is who he/she claims to be. Confidentiality guarantees that data cannot be read by unauthorized users. Privacy provides the users the ability to control who can access their data. Provenance allows an efficient tracking of the data and resources along with their ownership and utilization over the network. Integrity helps in verifying that the data has not been modified or altered. These services are currently managed by centralized controllers, for example, a certificate authority. Therefore, the services are prone to attacks on the centralized controller. On the other hand, blockchain is a secured and distributed ledger that can help resolve many of the problems with centralization. The objectives of this paper are to give insights on the use of security services for current applications, to highlight the state of the art techniques that are currently used to provide these services, to describe their challenges, and to discuss how the blockchain technology can resolve these challenges. Further, several blockchain-based approaches providing such security services are compared thoroughly. Challenges associated with using blockchain-based security services are also discussed to spur further research in this area.

## I. INTRODUCTION

A blockchain is a secured, shared and distributed ledger that facilitates the process of recording and tracking resources without the need of a centralized trusted authority. It allows two parties to communicate and exchange resources in a peer-to-peer network where distributed decisions are made by the majority rather than by a single centralized authority. It is provably secure against attackers who try to control the system by compromising the centralized controller. Resources can be tangible (e.g., money, houses, cars, lands) or intangible (e.g. copyrights, digital documents, and intellectual property rights). In general, anything that has a value can be tracked on a blockchain network to reduce its security risks and save the cost of security monitoring for all involved [1].

Recently, the blockchain technology has attracted tremendous interest from both academia and industry. The technology started with Bitcoin, a cryptocurrency that has reached a capitalization of 180 billion dollars as of January 2018 [2] [3]. According to the Gartner report in 2016, the blockchain technology is receiving billions of dollars in research and enterprise investments and much more is expected to come in the near future [4]. The technology currently spans several applications that are popular and driving the networking research. Such applications include healthcare [5], Internet of Things (IoT) [6] [7], and cloud storage [8]. Generally, the blockchain technology has proven its potential in any application that currently requires a centralized ledger. A practical example that employs blockchains is the Interbank Information Network provided by JP Morgan which provides fast, secured, and cheap international payments [9]. In addition, supply chain systems by IBM is exploring the potential of using blockchains in their services [10].

Among the blockchains' promising applications are network monitoring and security services including authentication, confidentiality, privacy, integrity, and provenance. Currently, these services are provided by trusted third-party brokers or using inefficient distributed approaches. As a result, security is a major challenge for current applications. On the other hand, the blockchain technology can provide security guarantees that resolve many traditional challenges in addition to providing a fully distributed, provably secure, and consensus solution. Fig. 1 illustrates the differences between the traditional and the blockchain-based access control. The same concept can be applied to the other security guarantees.

This survey focuses on the use of the blockchain technology to provide network security services and applications. We present the use of these services in the current applications, discuss the conventional techniques that provide these security services, and illustrate their challenges and problems. Then, we present how the blockchain technology can be used to resolve the associated challenges and highlight several proposed blockchain-based approaches that provide the desired security services. Finally, we discuss the current challenges faced with blockchain and some of the potential future research directions

This publication was made possible by the NPRP award [NPRP 8-634-1-131] from the Qatar National Research Fund (a member of The Qatar Foundation) and NSF grant CNS-1547380. The statements made herein are solely the responsibility of the author[s].

Tara Salman is with Washington University in Saint Louis, St. Louis, MO 63130 USA (email: tara.salman@wustl.edu)

Maeda Zolanvari is with Washington University in Saint Louis, St. Louis, MO 63130 USA (email: maede.zolanvari@wustl.edu)

Aiman Erbad is with Qatar University, Doha, Qatar (email: aerbad@qu.edu.qa)

Raj Jain is with Washington University in Saint Louis, St. Louis, MO 63130 USA (email: jain@wustl.edu)

Mohammed Samaka is with Qatar University, Doha, Qatar (email: samaka.m@qu.edu.qa)

in this field. It should be noted that the details of the blockchain technology and how it is used in other domains are out of the scope of this paper. We refer the readers to [1] and [2] for more details on the blockchain technology.

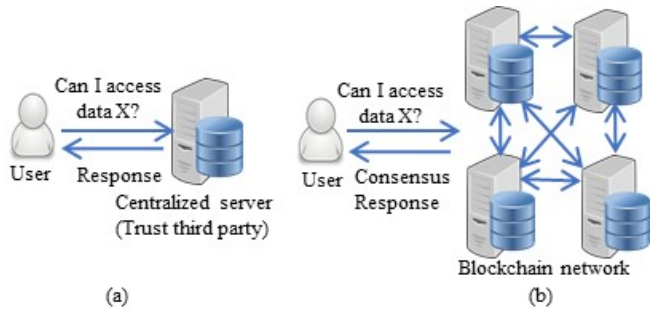

Fig. 1: (a) Traditional centralized access control guarantees (b) Blockchain-based access control guarantees

### A. Related Work

With the current growing interest in the blockchain technology, many new platforms and applications have been proposed. Several survey papers have been written to highlight the benefits of this technology for the current applications. Examples of such surveys include the blockchain technology for IoT [7], healthcare [5] and decentralized digital currencies [11]. Other surveys have discussed blockchain challenges, opportunities, and future visions. For example, the authors in [12] discuss the blockchain security issues and challenges. The work in [13] presents a thorough survey on blockchain security and privacy issues including possible attacks and countermeasures. Moreover, a recent special issue of IEEE spectrum is dedicated to blockchains and their potential uses [14].

This paper investigates the use of the blockchain technology in a different set of applications with rising interests that have not been discussed in the prior surveys. We aim to provide a comprehensive survey on the use of the blockchain technology in security services. The services can be offered by an enterprise and verified globally, offered by an enterprise but not verified, or presented as a research work. We strive these services to give insights on the current state-of-the-art technology and its challenges and discuss how the blockchain technology can be used to resolve these challenges.

### B. Security Services and Mechanisms

According to the X.800 family of standards [15], security services can be defined as the services that aid the open system interconnection protocols in providing adequate security to the transferred data over the system. These services can be divided into six categories: authentication, data privacy, data integrity, data confidentiality, non-repudiation and data provenance. The authentication service includes data origin authentication and entity authentication. The mechanisms to achieve this service include encryption and digital signature schemes. These mechanisms can be provided using public key cryptography, which will be explained later in Section III. The data privacy service can be achieved by access control mechanisms. The data confidentiality service can also be obtained by encryption and; therefore, public key cryptography can be used. The data integrity service can be achieved by message authentication codes using the secret key or the public key cryptography. The integrity mechanisms include replicating of the data and validating that replicas match. The non-repudiation service assures that no one can deny his/her action later and this can be provided using digital signature schemes; therefore, public key cryptography techniques can be employed. Further, we add the data provenance as another service to achieve tracking and monitoring of the data or resources. Table I summarizes these security services and their associated mechanisms.

In this paper, we consider the blockchain-based security services. Therefore, our discussion will include services such as authentication, data privacy, data integrity, and data confidentiality. Authentication and confidentiality are both provided by the public key cryptography; hence, these two will be combined in the same section. Privacy and integrity will be discussed in separate sections. It should be noted that non-repudiation is already provided by blockchain as will be explained later in Section II; therefore, we will not consider it among the services discussed later in the paper.

### C. Paper Organization

The rest of the paper is organized as follows: Section II gives a brief background on the blockchain architecture and its key properties and platforms. Section III discusses both the traditional and the blockchain-based approaches in providing authentication and encryption by public key cryptography and key management techniques. Section IV describes both the traditional and the blockchain-based approaches to provide privacy and access control lists (ACL). Section V presents both the traditional and the blockchain-based approaches to provide provenance services that track and report the data and resources shared in the network. Both the traditional and the blockchain-based approaches for integrity services to check for correctness and reliability of the data are discussed in Section VI. Section VII focuses on the challenges currently faced with the use of the blockchain technology and their effect on security services. Finally, Section VIII summarizes the discussion and highlights the main presented points.

TABLE I
SECURITY SERVICES VERSUS SECURITY MECHANISMS

| Services | Mechanisms | | | | | |
|---|---|---|---|---|---|---|
| | Encryption | Digital Signature | Message Authentication Code | Public key cryptography | Access Control | Provenance Techniques |
| Authentication | X | X | | X | | |
| Data Privacy | X | | | X | X | |
| Data Integrity | | | X | | | |
| Data Confidentiality | X | | | X | | |
| Non-Repudiation | | X | | X | | |
| Data Provenance | | | | | | X |

## II. BLOCKCHAIN BACKGROUND

In this section, a brief introduction to the blockchain technology is first presented. Following that, mining or block construction techniques are explained. The appealing characteristics of blockchains are also discussed along with a comparison of different open-source blockchain implementations. The objective of this section is to introduce the readers to the blockchain technology and its key principles.

### A. Blockchain Architecture

A blockchain consists of a database and a network of nodes, as illustrated in Fig. 2. A blockchain database is a shared, distributed, fault-tolerant and append-only database that maintains the records in blocks. Although the blocks are accessible by all the blockchain users, they cannot be deleted or altered by them. The blocks are connected to each other in a chain as each block has a hash value of its predecessor. Each block contains several verified transactions. Also, each block includes a timestamp indicating the creation time of that block, and a random number (nonce) for cryptographic operations. The blockchain network consists of nodes that maintain the blockchain in a peer-to-peer, distributed fashion. All nodes have access to the blocks, but they cannot completely control them.

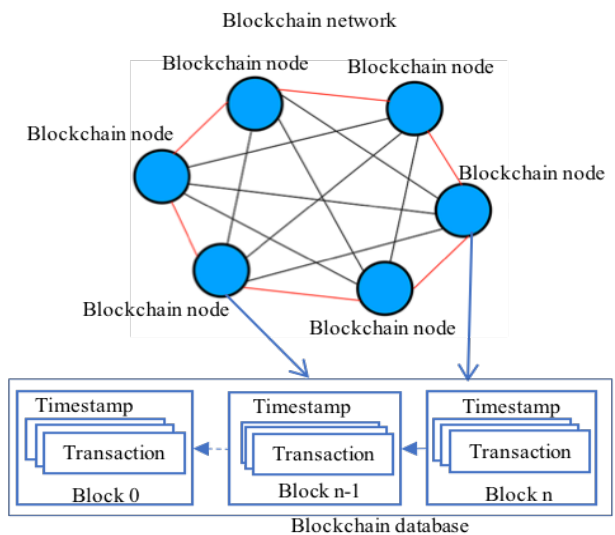

Fig. 2: Blockchain network, database, blocks, and transactions

The blockchain technology allows the communicating parties to interact in the absence of a trusted third-party. The interactions are recorded in the blockchain database providing the desired security requirements. When a blockchain user needs to interact with another user, it broadcasts its "transaction" to the blockchain network. Several nodes in the network check if the interactions are valid and construct a new block of valid transactions by mining (i.e., combining several valid transactions). The making of the blocks will be discussed further in the next subsection. If the new block is found valid, it is attached to the blockchain database and cannot be deleted or altered later. Otherwise, the block is dropped. Both the transactions and the blocks are signed; hence, they cannot be reverted or denied in the future.

The blockchain technology has three generations that support money transactions, assets, and smart contracts, respectively. The first generation was published by Satoshi Nakamoto in 2009 [1]. The application of this generation was restricted to money transactions and was implemented as a part of the Bitcoin cryptocurrency, which was the first application utilizing the blockchain concept. The second generation of the blockchain technology had broader use cases that exchanged assets rather than just money. In this generation, users own "shares" or "assets" and they can exchange any type of assets, including goods, properties and even votes [2]. In the third generation of the blockchain, smart contracts were introduced. A smart contract is a programmable contract that is checked by everyone in the network; thus, it compels both communication parties to strictly follow the contracts. The capabilities of blockchains were enhanced significantly within the third generation which led to its worldwide popularity and an increasing interest in its applications for several other critical services [6].

### B. Mining a Block in a Blockchain

Mining is the process of creating blocks that will be attached to the database. In some of the blockchain applications, such as in Bitcoin, the miner who creates the first valid block is rewarded. This reward is given by the system and is generally in terms of money for financial applications. Mining is one of the critical concepts in the blockchain technology. It allows nodes to create blocks which will be validated by others as well. If the new block is found as valid, it is attached to the blockchain database. Nodes that try to create blocks are called *"mining nodes."* The mining nodes race to validate the transactions and create a new block as fast as they can to win the reward.

Several approaches exist to decide which miner wins, including proof of work (PoW) [16], proof of Stake (PoS) [17], Proof of Space (PoSpace) [18], Proof of Importance (PoI) [19], Measure of Trust (MoT) [20], minimum block hash [21], and Practical Byzantine Fault Tolerance (PBFT) [22]. In the following, we summarize these major mining approaches (see also Table II).

- **Proof of Work:** PoW is the mining technique used in Bitcoin and is currently used by many other blockchain technologies. It requires the mining nodes to solve a hard-mathematical puzzle that is changed frequently and has been agreed by all the miners. Once a node validates the transactions and solves the puzzle, the block is submitted to the blockchain network. Other mining nodes validate the block to make sure that the submitter is not falsifying. Once it is agreed among the miners that the block is legit, it will be added to the blockchain and the submitter will be rewarded. The agreement here is based on a majority consensus. Thus, it is difficult to fake unless the attackers compromise more than 50 percent of the mining nodes. The problem with this approach is that high computational power is wasted in solving the mathematical puzzle [16].

TABLE II
COMPARISON OF DIFFERENT MINING TECHNIQUES

| Mining approach | Resources needed | Randomness | Implementations | Reward miner? |
|---|---|---|---|---|
| POW | High computation power | No randomness | Bitcoin | Yes |
| PoS | Wealth or stake | Randomized blockchain selection | Ethereum | No |
| PoSpace | High memory | No randomness | Permacoin | Yes |
| PoI | Node significance | No randomness | NEM | Yes |
| MoT | Trustworthiness | No randomness | Not implemented | Yes (trust) |
| Minimum block hash | None | Randomized blockchain selection | Bitcoin extension | Yes |
| PBFT | None | No randomness | Hyperledger | No |

- **Proof of Stake:** Unlike PoW, PoS does not require the mining nodes to solve a computationally expensive mathematical puzzle. Instead, the next block creator or miner is chosen in a pseudo-random way. The chance of a node being chosen to create the new block depends on the node's wealth or stake. In other words, the more money a node has, the higher its chances to mine a block. The native version of PoS does not award the miner; however, the extended versions award and punish the creators based on their performance. Selection based on the wealthiest account may result in a single account handling all the creations; hence, it may lead to an unfair distribution or even centralization. Therefore, a randomized node selection and a coin age-based selection have been proposed. In coin age-based method, the users that have not created any block for the past 30 days are considered for mining [17].
- **Proof of Space:** PoSpace is similar to PoW except that the puzzle requires a lot of storage. A miner proves its ability to create a new block by allocating the required storage space to perform mining. In other words, instead of having a high computational capability, the mining node needs to have a high storage capability. Several theoretical and practical implementations of PoSpace have been released; however, the required high memory space is a challenge similar to the computation challenge of PoW [18].
- **Proof of Importance:** PoI is a mining technique that calculates the significance of an individual node based on the transaction amount and the balance of that node. It assigns a priority with a hash calculation to the more significant nodes. Further, the node with the highest priority is chosen for the next block creation [19].

- **Measure of Trust:** Another way to perform mining is to use dynamic trust measurements and select the node with the highest trust level as the block initiator [20]. The trustworthiness is based on the nodes' behaviors; therefore, good behaving nodes that follow the protocols are rewarded. More specifically, the trustworthiness could be formulated as the expected value of the node's behavior in the future. This, the trustworthiness is approximated by the history of good and bad actions that the node has taken so far. The MoT approach could be subject to malicious attacks if a specific node plans to increase its trustworthiness for several iterations in order to attack the network later. The authors in [20] proposed several mechanisms to handle such attacks.
- **Minimum Block Hash:** In [21], the authors proposed an approach for mining where the miner is chosen randomly and not based on its resources. The system selects the miners based on a generated minimum hash value across the entire network. Thus, the selection of the next miner is randomized and the probability of selecting the same miner is low. This approach was implemented on a modified Bitcoin network and it was shown to offer energy savings for mining. However, it has not been adopted by the Bitcoin community.
- **Practical Byzantine Fault Tolerance:** Unlike others, PBFT [22] is a consensus approach that does not include any type of resources but utilizes the blockchain consensus based on the Byzantine fault tolerance approach. In this approach, first, a leader is selected and agreed among the nodes. The leader decides on the transactions' validation and publishes a block to all the nodes in the blockchain network. A transaction is committed to a new block only if two-thirds of the mining nodes verify its correctness. The leader changes frequently; therefore, the approach is not considered as centralized. PBFT has been shown to be faster than other methods; however, it suffers from scalability issues due to the resulting communication overhead as discussed in [23].

*C. Key Properties of Blockchains*

Key properties of the blockchain technology include their distributed nature, decentralized consensus, trustless system, cryptographic security, and non-repudiation guarantees. In Table III, we briefly summarize these properties and the problems they try to solve.

TABLE III
KEY PROPERTIES OF BLOCKHAINS

| Property | Problem to be solved | Blockchains' solution |
|---|---|---|
| Distributed Nature | Current applications are distributed by nature, therefore, require distributed control and security mechanisms. Most of the current practical security solutions are centralized and inefficient for these applications. | The blockchains are distributed by nature. Thus, blockchain-based security services can be implemented in a distributed fashion |
| Decentralized Consensus | Centralized decisions by one controller can make the controller a single point of failure. | The blockchain decisions are achieved by decentralized consensus, majority votes, and nodes agreement. |
| Trustless System | Security provided by third parties can impose security and privacy risks if the party is compromised | The blockchain technology imposes a trust of majority votes, which is impenetrable to compromise unless attackers have control over the entire system. |
| Cryptographic Security | Algorithms for security should prove that they are supremely difficult to break. | The blockchains use elliptic curve cryptography that is difficult to break. Further, the trustless system and the decentralized consensus make it even more difficult to break. |
| Non-repudiation Guarantee | Users can deny their interactions in the system | The blockchains use signatures of transactions and blocks in addition to permanent databases such that transactions cannot be denied later. |

### D. Blockchain Open-Source Implementations

As there are many open-source implementations of the blockchain technology, the choice of which implementation to use is challenging. In Table IV, we compare different aspects of several popular blockchain implementations. We will be referring to these implementations throughout this paper when we discuss the blockchain-based security services. It is important to keep these features in mind to highlight the properties of each implementation. It should be noted that these are not the only implementations and many others exist in the literature. However, these are the most popular ones used in the majority of the blockchain applications.

TABLE IV
OPEN-SOURCE IMPLEMENTATIONS OF BLOCKCHAINS COMPARISON

| Platform | Smart contract | Mining | Advantages | Disadvantages |
|---|---|---|---|---|
| Bitcoin [25] | No | POW | • Scalable in terms of the number of nodes and users. <br> • Currently most popular | • Computationally expensive. <br> • Time consuming |
| Ethereum [26] | Yes | POW or PoS | • Scalable <br> • For PoS mining, no computation is required. | • Require stake or wealth to be selected for mining |
| HyperLedger [27] | Yes | PBFT | • No minting and thus faster than all others (promised) | • Scalability problem. Does not scale above 20 nodes as reported in [24] |

### E. Summary

A blockchain is a distributed, shared, append-only, and permanent database that was first utilized by Bitcoin for cryptocurrency applications. Its key properties include distributed nature, consensus, trustless system, cryptographic security and non-repudiation guarantees. These properties make the blockchain technology a potential approach for the current distributed applications including IoT, healthcare, and automated supply chains. Several variations of the blockchain technology exist in the literature to solve the challenges introduced in the first generation. One of the critical challenges in Bitcoin mining is the computational capability that is required to perform mining. Alternatives to PoW mining include: PoS, PoSpace, PoI, MoT, minimum block hash, and BPFT. These alternatives resulted in many open-source blockchain platforms that developers can choose depending on the application. In the remainder of this paper, we assume that the reader has the knowledge of the discussed platforms and their variations, as well as the advantages, and disadvantages of each.

### III. ENCRYPTION AND THE AUTHENTICATION SERVICES

Encryption and authentication are two of the most important security services that must be provided in any network system. In general, these services can be granted using public key cryptography as one of the well-known security frameworks. The public key cryptography techniques require the entities to have private and public information. They need an infrastructure to create, revoke, manage, distribute, use, and store the generated keys or the generated information. In this section, the public key cryptography and its uses in today's applications are first discussed. Following that, an introduction to the public key management techniques and their challenges are presented. Then, an overview of how the blockchains can be used to solve these challenges and some blockchain-based key management techniques are discussed and compared.

### A. Public Key Cryptography and Its Services

Public key cryptography, also known as asymmetric cryptography, is a cryptographic technique that uses a pair of keys: public keys which are distributed over the system and private keys which are kept secret. It was introduced initially by Diffie and Hellman in 1975 and is still widely adopted. The basic idea is to use one of the keys to do a task (encryption or signature) and use the other key to do the reverse of that task (decryption or validation). In this way, every entity can verify the message coming from a certain user by the user's public key. The reply message can also be encrypted before sending it back. Only that specific user can sign/decrypt the message with its private key.

The public key cryptography can be used for many security services including the entity authentication and the confidentiality. As illustrated in Fig. 3, the entity authentication service can be provided by the signature/verification procedure. An entity sends a message signed with its private key and everyone can verify/authenticate that entity by validating the signature with the entity's public key. Since the private key is kept confidential, no one can sign the message except the entity itself or someone who has access to the private key. On the other hand, the verification is done with the public keys. Thus, everyone with the user's public information can verify and authenticate that user.

The confidentiality service can be achieved by encryption/decryption, which is a similar procedure. The encryption is done by the sender with the receiver's public key. The decryption is done by the receiver with his private key. Only the receiver, or someone who has the receiver's private key, will be able to decrypt and understand the data. Therefore, the confidentiality is guaranteed.

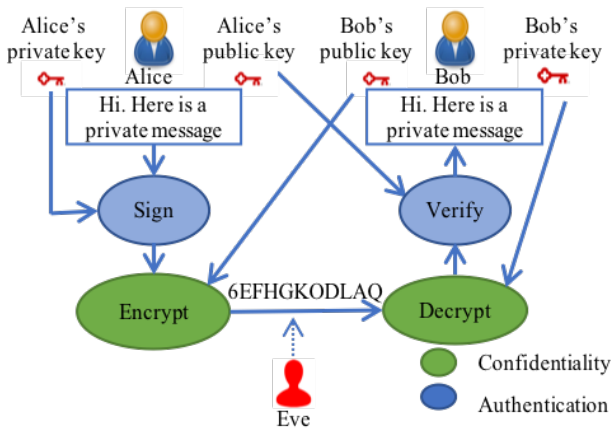

Fig. 3: Illustration of public key services

### B. Services' Importance for the Current Applications

Entity authentication and message confidentiality are the most critical services in almost all of the current network applications. A smart healthcare environment is a typical example of the importance of these services. The system is required to secure the transmitted data in order to keep patients' privacy from intruders. Further, it is crucial to authenticate the right doctor, the hospital and the pharmacy and secure their access to the data.

To generate the private/public keys for the system, many algorithms have been proposed, including RSA [28], ElGamal [29], and elliptic curve [30]. Discussing these algorithms is out of the scope of this paper. However, in general, these are complex and need an infrastructure to generate and manage the public/private keys. The certificate authorities (CA), the web of trust (WoT) and the entity-based cryptosystem have been introduced to create, manage, use, store, and distribute the keys. In the following subsections, we discuss both the traditional and the blockchain-based key management approaches, including CA and the web of trust. In a later subsection, we discuss the entity-based cryptosystem which is the current trend that extends the CA mechanisms to make a better use of the public key cryptography.

### C. Key Management by the Public Key Infrastructure (PKI)

The public key infrastructure (PKI) is one way to provide the key management for the public key cryptography. Traditionally, there are two conventional approaches to achieve PKI, centralized by a CA and decentralized by WoT. The CA-based PKI is the most commonly used approach and it has been standardized in the X.509 standard [31]. In this approach, the CA is a third-party entity that is trusted by all members in the system. The CA issues "certificates," which authenticate users and bind each user to a public key. A signed certificate, binding a user to its public key, will authenticate the ownership of that public key to that specific user. The other traditional approach is WoT, which was proposed in 1992 by Phil Zimmerman. This technique utilizes a decentralized approach in which the keys are generated locally and will be trusted if they are verified by at least one other trusted user in the system [32].

### D. Problems with the Traditional PKI Systems

Both of the traditional techniques suffer from several challenges which are discussed in this subsection.

The CA-based PKI comes with three major challenges: a trusted third party, a single-point-of-failure, and cost. The users of the systems must trust the CA in generating and managing their public keys which imposes high-security risks if the CA is compromised. This architecture has a single point of failure as the whole system fails if the CA fails. Furthermore, the management of the public keys by one centralized CA can be both expensive and inefficient, especially with the current massively distributed applications where a large number of users are involved [33].

On the other hand, in the WoT-based PKI, the signers need to build trustworthiness. The users join the network only if they are trusted by another "trusted" member. In other words, new members joining the network need to build prior trust with other members who are already in the system. This can lead to a barrier for new members entering the network [33].

Moreover, both the CA-based and the WoT-based PKI are unable to provide identity retention. That is, it is possible for a user to impersonate the identity or the public key of an already registered user. Some proposals have been offered to solve this problem; however, they are mostly log-based, which could be highly complex, especially in the case of the worldwide distribution of the users [33].

### E. Blockchain-Based PKI Concept

The distributed, the event-recording and non-reproducibility features of the blockchain technology make it a desirable technique for several applications. Particularly, these properties prove the blockchains' suitability for PKI and domain name services (DNS). Since the blockchain-based PKI solutions are distributed; they have no centralized point of failure. The trust is built based on the majority vote of the miners; hence, there is no single trusted third-party and it does not require prior trustworthiness in the system. More importantly, the blockchain technology has several open-source implementations, which helps build cost-effective and efficient solutions. The problems with the traditional approaches and how the blockchains can solve them is summarized in Table V.

In the following, we discuss several approaches to achieve blockchain-based PKI.

#### 1) Instant Karma PKI (IKP)

The Instant Karma PKI (IKP) framework extends the traditional CA approach by recording the CA behavior to the blockchain database. In this way, misbehaving or compromised CAs can be detected by the network and a riposte must happen.

The event recording feature of the blockchains facilitates the CA tracking and monitoring by the blockchain users and helps detect the misbehaving CAs. This approach can reduce the trust problem in the traditional CA-based algorithm as eventually misbehaving CA can be detected.

TABLE V
THE TRADITIONAL PKI PROBLEMS AND THE BLOCKCHAIN-BASED SOLUTIONS

| The traditional approach | The problem | The blockchains solution |
| --- | --- | --- |
| CA-based PKI | Third party trust<br>Single point of failure<br>Cost of deployment | The distributed consensus property of blockchains<br>No centralized authority<br>Open source implementations |
| WoT-based PKI | Prior trustworthiness | Does not require any previous trust. The decisions are made based on majority votes |
| CA-based and WoT-based PKI | Identity retention | The technology has an event recording database and it thus can verify if the public key has been registered before or not |

IKP is a research work that was proposed in 2017 and verified in terms of cost saving and distribution. An open source implementation was also promised but has not come available at the time of this writing [34]. However, having a CA in the system still lead to a single point of failure system. Trying to solve this matter by having several CAs imposes cost; thus, leading to even a more expensive solution.

*2) Pemcor*

Pemcor utilizes the blockchain database as a distributed and secure data store [35]. The idea is to let the CA issue a certificate which is not signed. Instead, the hash value of the certificate is stored in the blockchain which is controlled by authorities, like by banks or governments. Such authorities share two blockchain databases, one for the generated certificates and one for the revoked certificates. When verifying, the authority checks its maintained blockchain data stores. If the hash of the certificate exists in the generated certificate blockchain and is not in the revoked certificates blockchain, the certificate is valid; otherwise, it is not. This idea is simple and provides several advantages such as an easy verification with low delay guarantees.

Pemcor is part of a project that aims to find a solution for identity proofing and replaces the traditional knowledge-based verification. The project was proposed and documented theoretically by several white papers in 2016, however, it still lacks the complete implementation and the evaluation of the system. Given other approaches presented in this section, this work is not expected to contribute further to the blockchain-based PKI systems.

*3) Gan's Approach*

In [36], the authors propose a key-based authentication system dedicated to the IoT environments. The idea is to use a private blockchain for storing the nodes' latest public keys, validating the keys, and allowing others to request the nodes' keys. The architecture of this approach is illustrated in Fig. 4, where a Centralized CA (CCA) is assumed to be fully secured. Several validators, donated as Device Manufacturer Validators (DMVs), are connected to the CCA.

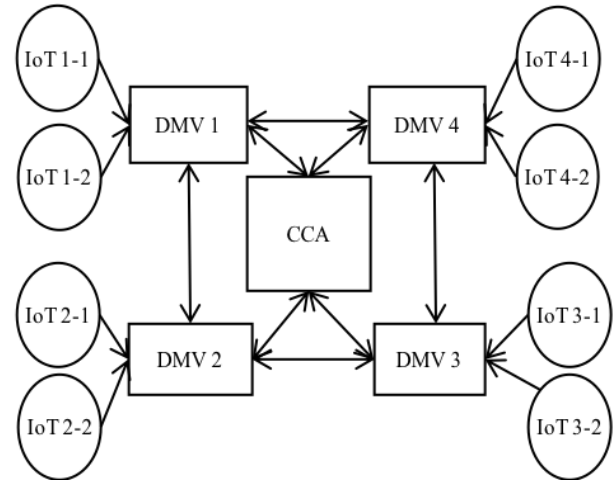

Fig. 4: The architecture of Gan's approach

The DMVs are hosted by the IoT manufacturers and they are required to have the computational capabilities to generate the public/private keys, to perform mining and to maintain the blockchain database. The IoT devices are connected to these validators and are assumed to be simple without any computational capability. Initially, a DMV joins the blockchain network by requesting the CCA to authenticate it. The CCA validates the DMV and constructs a transaction that contains the DMV public key, the validator address, and the CCA's signature. The transaction is submitted to the blockchain and the DMV is now known to the others. Accordingly, the DMV can add a new IoT node by submitting a transaction containing the node's public key and address to the blockchain. Furthermore, the DMVs can update or revoke their IoT devices' public keys by submitting transactions.

This work has an open-source implementation that is available on GitHub and referenced in [36]. The implementation utilizes Network Simulator version 3 (NS3) to build and evaluate their proposed approach. Even though the approach was initially applied to the IoT platforms, the idea can be implemented in any other networking applications including sensor networks, health care, or even micro clouds platforms.

It should be noted that these previously discussed three approaches use the blockchains only as distributed databases to share and validate the keys. In other words, they do not benefit from the other important properties of the blockchains including the distributed consensus and the non-repudiation guarantees. In addition, as it was mentioned before, having a CA that generates the keys does not resolve the problems of centralization; hence, the single point of failure in the traditional PKI approaches.

*4) Distributed PKI (DPKI)*

In [37], the authors sketch the principles of an appropriate blockchain-based PKI, which is referred to as Distributed PKI (DPKI). The DPKI uses the blockchain technology as a distributed, trustless database that eliminates the need for a CA and gives the users the direct control and ownership of their data. This work uses a web registration domain, where the user spawns its public/private key and submits the public key to the blockchain network as a transaction. In this work, it is claimed that the blockchain technology can resolve the traditional problems and protects the network against man in the middle attacks. This protection is granted by linking the most recent key of the user to his/her identity.

The paper did not include any implementation-related aspects; nevertheless, it introduced the possibility of blockchain-based PKI, which was later implemented in many other works as will be discussed next.

*5) Blockstack*

Blockstack ID is an appropriate blockchain-based approach that uses Namecoin to build a distributed PKI system. Namecoin, [39], is a fork of Bitcoin that allows data storage within the blockchain transactions. It is implemented by defining a name-value pair that is used to store usernames and can be recorded in the transactions. Namecoin was originated to store the DNS names, allowing users to register their human-readable name and associating names with the corresponding public keys.

Blockstack ID modifies Namecoin by adding another name-value pair dedicated for the public keys. The advantage of using Namecoin is that it already supports the name-value pairs in its transactions. Thus, the public key is the value and the name is the identity of the owner. Blockstack implementation binds the user identity to an elliptic curve public key which is one of the strongest public key cryptography mechanisms to date.

Blockstack was released as an open-source software in 2014 and is currently serving as a PKI system for 55,000 users. It is probably the most popular blockchain-based PKI among other techniques discussed in this section. However, some issues such as how the system would handle the public key updates, the lookups, and the revocations have not been considered in Blockstack. Also, the identity retention problem is not been resolved.

*6) Certcoin*

Certcoin, [40], is another fully decentralized PKI that relies on Namecoin to build its platform. Unlike Blockstack ID, this platform provides the identity retention guarantee. As in the traditional PKI approaches, this system is composed of 5 functions: registration, update, lookup, verification and revocation. During the registration, the owner originates its own private and public keys locally. It keeps the private key to itself and submits a transaction of the public key and its signature to the blockchain. The blockchain network verifies the transaction signature and the fact that this ownership was not registered before in the system. If the verification is successful, the (ID-public key) tuple is added to the blockchain; otherwise, it is dropped. To update the public key, the owner submits a transaction containing the identity along with the previous public key, the new public key, and the signature. Miners need to verify that the signature is correct, the identity exists in the blockchain, and it is associated with the previous public key. Then the mined blocks are broadcasted to the network to be verified. The verification follows a similar process where the owner submits a transaction requesting the blockchain network to verify the key which can be done by the miners and other blockchain nodes.

Certcoin has 3 versions, numbered 0, 1, and 2. The first version (version 0) required complex computations and operations while the second and the third versions tried to reduce this complexity by accelerating the blockchain processing. Version 0 had all the five functions submitted and mined by the blockchain network as in Bitcoin. However, those functions are complex and result in a computationally expensive process. Thus, versions 1 and 2 tried to reduce this complexity by a cryptographic accumulator and a distributed hash table, respectively. A cryptographic accumulator is a space-efficient data structure that is used to reduce the time and the complexity of the verification process [41]. A distributed hash table supports fast look-ups for the public key queries; hence, the complexity of the lookup and the verification is reduced [42]. Therefore, the verification, the lookup and the update functions have been simplified in both versions 1 and 2.

Similar to Blockstack, Certcoin is an open-source implementation that was first released in 2014. The project was one of the first blockchain-based PKI and is offered by MIT [40]. However, Certcoin is less popular compared to Blockstack due to the lack of proper documentations and the lack of updates to the software.

*7) Guardtime Solution*

Guardtime provides another solution for secure authentications of the IoT devices using the blockchains and physically unclonable functions (PUFs). A PUF is a digital fingerprint hardware that serves as a unique identifier of the devices. The PUFs use the unique characteristics of each device to generate its unique private/public keys. Guardtime employs PUFs to generate the public/private keys. The public keys are submitted to the blockchain in transactions. IoT devices have limited memory; hence, they cannot store large private/public keys [43]. In other words, the solution provided by PUF and Guardtime helps such devices regenerating the same key each time it is needed.

Guardtime is an enterprise that currently offers blockchain-based solutions for several industries including the insurance companies, the physical supply chains, the cloud providers, and many others. Guardtime Federal is a fork from Guardtime that started in 2014 and is dedicated to providing cyber-security solutions for the US department of Defense, the U.S. Intelligence Community, and other U.S. Government departments. All solutions offer the confidentiality and authentication services. However, these solutions are mainly dedicated to supply chains and their integrity assurance, as will be discussed in Section VI-E.

*8) Blockchain-Based Trust and Authentication for Decentralized Sensor Networks*

Moinet at al., [44], propose an approach of using the blockchain technology as a database to store the public keys, the digital signatures and some peers' information in a wireless sensor network. This approach is similar to Certcoin, since it allows the nodes to verify and authenticate each other using the blockchain network. Initially, when a node wants to join the network, it submits a credential transaction, or a credential payload as referred in the paper. This transaction has the master public key and a signed hash value that is used to authenticate the node. A node can submit a transaction to renew or revoke its own public key. In addition, this approach introduces a "blame transaction" which defines the trust level of all nodes in the network. A node is blamed whenever its trustworthiness goes below a certain defined level. The blame transaction includes the node that generated the blame, the blamed node and the block that had the node ID and the public key included. Furthermore, all the blocks must have a miner's approval transaction to be valid. A miner's approval transaction would include the miner ID, a nonce, the new node's public key, and the signature of the miner.

It should be noted that this approach is an extension of the web of trust traditional approach, where a node can join the network only if another node (miner) approves it. Also, this is mainly a research work that does not include any proper blockchain implementation, thus, is not getting practical and popular compared to Blockstack or Certcoin.

### F. Identity-Based Cryptography (IBC)

Recently, identity-based cryptography (IBC) has gained interest in the network security community. IBC is a public key mechanism that uses the node's ID as the public key rather than generating the traditional lengthy public keys. A node's ID can be the node's name or any arbitrary string that can be used as the public key. The encryption approach, as depicted in Fig. 5, consists of four phases: setup, extract, encrypt, and decrypt.

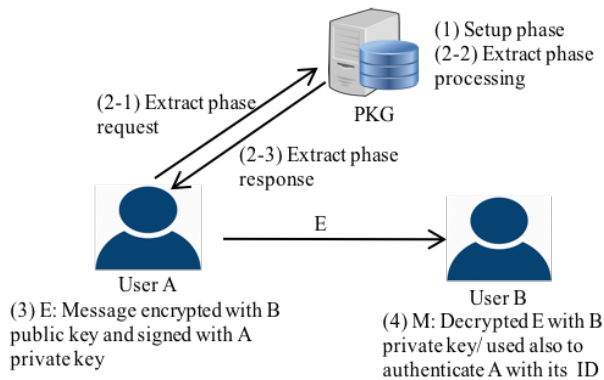

Fig. 5: Identify-based encryption phases

In the setup phase, a private key generator (PKG) generates a master secret key along with some public system parameters. The secret key is kept private while the system parameters are made public. To extract the keys, the generator uses the system parameter in addition to its master secret key and the user's ID. These parameters are used to construct a secret key which is sent back to the user. For other nodes to encrypt a message, they use the ID and the public parameters to generate a ciphertext. The user uses its own private key to decrypt the message. A similar approach is used for the signature and the verification, where the signature is generated with the node's secret key while the verification is done with the node's ID and the public system parameters [45].

A generalization of the IBC is to build a Hierarchal IBC (HIBC) where the public key has a hierarchal identity basis that can be represented by a tree. For example, the public key of Alice at organization X is Alice@X rather than Alice. The encryption phases of HIBC are the same as the four phases in the IBC with an additional phase, called delegation. This phase allows an entity to generate secret keys for its children. In other words, the system needs one PKG to generate the secret key for the root. Others secret key can be driven from that key [46].

### G. Problems with the Current IBC-Based Approaches

The problem with both the IBC and the HIBC approaches is that they require the PKG to generate the private keys. Therefore, the system is centralized which makes it a single point of failure and imposes a third-party trust requirement. It is centralized as the PKG is the only authority that can generate key pairs. If the PKG is compromised, the whole system is compromised. In other words, the IBC and the HIBC have the same limitation as the CA-based PKI traditional approach. Moreover, the PKG generates the users' private keys; hence, all the users should trust the PKG not to misuse their private keys.

### H. Blockchain-Based IBC

Similar to the blockchain-based PKI approaches, the blockchain technology can be used as a distributed database to resolve the problems of the traditional IBC approaches. Since, the blockchain technology has a decentralized database, it solves the problems of centralization and the single point of failure. It does not require a third-party trust as the users can generate their own master keys. However, if a user has limited resources and cannot generate its own key, it can delegate that to any other node that it trusts.

The basis of the blockchain-based IBC systems is to let the users generate their own keys. This indicates that the setup and the extract phases are done at the user level. Then, the public parameters are submitted to the blockchain as a transaction. The blockchain nodes check whether these parameters are valid and they have not been used before. Any user can later query the blockchain network for other users' public parameters which are used to authenticate the user and encrypt the confidential messages.

#### 1) Blockchain-Based IBE for Information-Centric Networking (ICN)

The blockchain-based IBC can be applied to secure the Information-Centric Networking (ICN) which considers "content names" as the main element for security, i.e., as the basis for inter-network communication. Therefore, it is practical to utilize the HIBC approach to secure the ICN, as the contents are designed to be hierarchical [46]. In [47], the authors utilize the blockchain-based IBC to provide distributed security for the ICN networks, where a content owner wishes to share some data with the subscribers. This approach consists of two phases: setup and retrieval. In the setup phase, the owner generates the public parameters and a secret key that are required by the HIBC. In other words, the owner acts as a PKG for itself. It registers the public parameters to the blockchain network by submitting a new transaction. Then, when a subscriber wants to access the data, it queries the network for the public parameters of the corresponding content. This means that instead of consulting the centralized PKG in the traditional HIBC algorithm, the subscriber consults the distributed blockchain. These activities are recorded in the transactions; hence, cannot be denied once the transactions are committed to the blockchain database. Thus, the blockchain technology can provide both the integrity and the provenance services, in addition to solving the centralized architecture challenges.

TABLE VI
BLOCKCHAIN-BASED PKI APPROACHES COMPARSION

| Approach | Blockchain platform | Modifies implementation? | Public key generation | Corresponding traditional approach | Main utilized blockchain properties |
|---|---|---|---|---|---|
| IKP | Ethereum | No | By CA | CA | Distributed Database |
| Pomcor | Ethereum | No | By CA | CA | Distributed Database |
| Gan's [36] | Their own | Yes | By CA and DMV | CA | Distributed Database |
| DPKI | - | - | By user | WoT | All |
| Blockstack | Namecoin (Bitcoin) | No | By user | CA and WoT | All |
| Certcoin | Namecoin (Bitcoin) | Yes | By user | CA and WoT | All |
| Guardtime | - | - | By user (PUF) | CA and WoT | Distributed Database |
| [44] | - | - | By user | WoT | Distributed Database |
| [47] | Namecoin (Bitcoin) | Yes | By user | IBC | All |

This scheme can be applied to any ICN or other similar applications. However, the initial implementations of this scheme showed a high level of complexity in generating the public keys. This problem resulted in restricting the scheme's practicality and popularity among the resource-limited ICN applications that are emerging.

*I. Summary*

The public key cryptography is an important security framework that is used widely to provide the authentication and the confidentiality services. Such services are critical for most of the current applications including IoT and healthcare. A management system is required to provide a proper infrastructure for such services. PKI is a framework to generate, distribute and manage public keys for the entities in the system. In this section, we discussed the traditional PKI systems and how the blockchains can be used to resolve their problems. We further presented several proposed approaches providing blockchain-based PKI solutions. Furthermore, we discussed the IBC technique which is a recent popular technique in providing security services. We presented the traditional IBC, their problems, the blockchain-based IBC, and a proposed approach for a blockchain-based IBC. Table VI shows the comparisons among the discussed approaches from different perspectives. It should be noted that even though these systems exist and are open-source, only few are utilized in real-world applications.

## IV. PRIVACY SERVICES

A privacy service offers the user the rights to control and set rules for its data and resources accessed by the network. In other words, it enables the data or resource owners to control the disclosure of their information. This is generally done by letting the user define his access control list (ACL). In this section, we investigate the requirements of providing the data privacy, its importance for the current applications, the traditional techniques for privacy, and the challenges currently faced in providing the privacy service efficiently. Then, we give an overview on how the blockchains can be used to provide privacy and summarize a few existing blockchain-based privacy providing systems.

*A. Data Privacy and ACL*

The data privacy requires that all personal and sensitive information remain confidential (not public) and access to them can be controlled by the data owners. The ACL assures that by defining a set of rules stating who can access a specific set of data and when. To illustrate the privacy problem, consider the users in organizations such as Facebook, Google, banks and government surveillance. Each user must provide his/her personal information. Thus, these organizations have a massive amount of personal data that should not be made public. Individuals have little or no control over the storage and the access to their information. Therefore, the data privacy can be violated. Many controversial incidents have been reported, especially with banks and government surveillance [48] [49].

The privacy concerns exist whenever the data is collected, stored, used, destroyed, or even deleted. In other words, privacy applies to the data in motion and at rest. Several federal laws have been developed to prevent information leakage; as an instance, the healthcare information privacy laws [50]. For all these reasons, privacy is a major concern for application and network developers.

*B. Importance of Privacy in Current Applications*

The data privacy is a prominent interest in the era of cloud computing and networking systems where many users share the same physical storage or network. Application developers migrate their storage and computations to the clouds and require the data privacy to be granted. Moreover, IoT, healthcare, smart grids, and several other popular networking applications need to process and store a massively large amount of data, generally using cloud computing. Privacy is a critical requirement for most of these applications that are involved with personal information or location knowledge. The problem of privacy is intesified in case of using multiple clouds and internetworking among them.

### C. Traditional Techniques for Data Privacy

Generally, the data privacy can be provided by delegating the ACL definitions to the data owners and using encryption techniques to prevent others from accessing the data. Hence, the organizations who amass or process the data have no rights to access the if the ACL does not permit. Design and implementation techniques to provide the privacy service is among the most active research topics, and several techniques have been proposed so far. For example, homomorphic encryption, which allows the computation and the processing the encrypted data and returns encrypted results, is one way to provide the data privacy service [51].

Another privacy aspect, which is out of the scope of this paper, is hiding the user's identity. Data anonymization and differential privacy mechanisms hide the identity of the user and make it difficult to link the data to its owner. For example, K-anonymity, a common way to anonymize the datasets, requires the sensitive information to be similar to at least K-1 other records [52]. L-diversity, an extension of the K-anonymity approach, guarantees that the sensitive information is stored in "diverse enough" possible locations [53]. T-closeness is another approach that looks at the distribution of sensitive data [54]. Differential privacy uses data perturbation techniques or adds noise to them before sharing the data [55]. Most blockchain implementations provide pseudo-anonymous user privacy. For example, Bitcoin utilizes the hashes to identify the users, rather than their real names. The users stay hidden from others and remain anonymous to the system unless sophisticated attack actions are taken [1].

### D. Problems with the Traditional Techniques

Despite the fact that several research efforts exist, provide an efficient data privacy service is still challenging. Some of the challenges include efficiency, scalability, data ownership and lack of systematic data lifecycle approach. In the following, we briefly summarize these problems and refer the readers to [56] for more detailed discussion.

- **Efficiency and Scalability:** Most of the data privacy techniques rely on complex cryptographic algorithms; hence, they are inefficient and difficult to scale with large applications. Recent research tries to reduce the complexity and enhance the efficiency of these cryptographic techniques [57]. However, the proposed approaches still lack practicality in most cases. Further, most algorithms fail to scale with the massive amount of data processing required in the current networks.
- **Data Ownership and Control:** The questions of who owns the data and who can modify it are critical in privacy. The owner generally is the party that decides the access control rules for the data. Unfortunately, the traditional techniques discussed in the previous subsection still lack an answer to the ownership question.
- **Systematic Data Lifecycle Approach:** A framework for the data privacy needs to be constructed to systematically define the lifecycle of the data. This framework should identify the phases, define their privacy requirements, and allow flexibility in the lifecycle changes. These phases can include the acquisition, the sharing and the deletion of the data and the resources involved in the system . However, a systematic approach is still missing in most of the proposed privacy techniques.

### E. Blockchain-Based Data Privacy Techniques

The blockchain technology can be used to provide decentralized end-to-end data privacy guarantees that can resolve some of the problems discussed in the previous subsection. Specifically, it can provide the data ownership solutions and dynamically change the access rights when needed. However, since the blockchains depend on cryptographic techniques, the blockchain-based techniques are still complex. The problems associated with the traditional approaches and how the blockchain technology can solve them are presented in Table VII.

TABLE VII
TRADITIONAL DATA PRIVACY PROBLEMS AND BLOCKCHAIN-BASED SOLUTIONS

| The traditional approach | The problem | The Blockchain solution |
|---|---|---|
| ACL Definition and Monitoring in addition to encryption (Data Privacy) | Efficiency | No solution by the blockchains. The homomorphic encryption may be still used and thus the problem still exists. |
| | Scalability | Generally, the blockchains scale better than the traditional approaches, however scalability is still a challenge in the blockchains, which will be discussed in Section VI |
| | Data ownership and control | The blockchain technology can resolve the problem as it can record the ownership and the changes in the data. The user has full control to define its ACL. |
| | Systematic lifecycle | The blockchain users can update their smart contract or establish new contracts easily and thus the changes can be flexible. However, they are still under user control. A programmable ACL should be written over the blockchains to make the approach systematic. |
| Anonymity and Differential Privacy (User Privacy) | Complexity | Some blockchain platforms can provide a pseudo-anonymous user privacy. For example, Bitcoin utilizes the hashes to identify the users rather than usernames. The users are hidden from other nodes and remain anonymous to the system unless further actions are taken. Using this, the pseudo-anonymity is provided. |

The idea behind the ideal blockchain-based data privacy is to build a blockchain layer over the data storage layer, let the owner define the desired ACL through smart contracts, and publish the ACL and the data to as the blockchain transactions (encrypted using sophisticated encryption techniques). In this way, organizations such as Facebook or Google will not own the data as happens in the traditional techniques. However, they will be a part of the blockchain network and they will be able to process the data only when the ACL allows them. This type of blockchains is called the *permissioned blockchains*. Policies to define the data access are either based on the smart contracts or on the data management messages. Further, an off-chain database can be used to store the encrypted data as the blockchain memory is limited and cannot store massive amounts of data. In the following, we discuss several recent approaches that utilize the blockchain technology to provide the privacy service.

*1) Zyskind's Approach*

Zyskind, Nathan, and Pentland, [20], propose a decentralized data privacy approach that ensures the users' control over their data and uses the blockchain blocks to store the data and the ACL. As illustrated in Fig. 6, the system is composed of three main components: users, providers and the blockchain network. The users are nodes interested in downloading an application or using a service. Providers, who hold such services or applications, need to process the users' personal data for operational and business purposes. The blockchain nodes are the untrusted entities that constitute the blockchain network and have a distributed data store (off-chain data store). The data is distributed and replicated among the data stores to ensure the privacy and the high availability services.

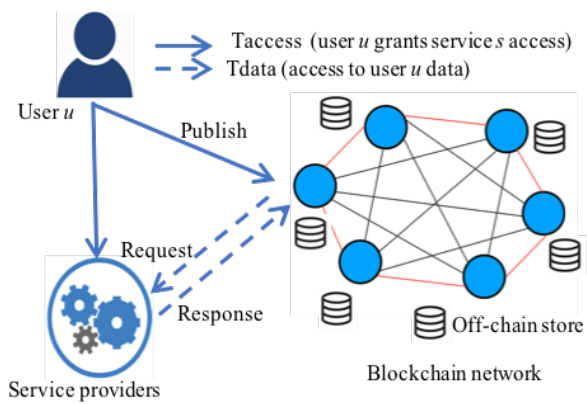

Fig. 6: Zyskind's proposed system components

The blockchain network accepts two types of transactions: $T_{access}$ and $T_{data}$. The $T_{access}$ is used for the control and the management operations on the data, such as defining the ACL and modifying the access rights. The $T_{data}$ is used for data storage and retrieval. The owner can change the permission and the access controls by sending a policy set in the $T_{access}$ transaction, which is checked for correctness by the blockchain nodes. Similarly, a user or a service provider can access the data by sending a $T_{data}$ transaction which will be approved by the blockchain nodes if the policies specified earlier are met. The returned response (access information or denial) is encrypted; hence, unauthorized users cannot have access the data.

Zyskind's Approach is a research work that has been verified theoretically and practically in their paper [20]. However, the open-source implementation of the proposed approach is still missing thus, practicality and suitability of the proposed approach are still to be testified.

*2) Blockchain-Based Data Sharing (BBDS)*

Blockchain-based data sharing (BBDS) is another approach proposed to provide privacy for the medical records in a cloud environment [58]. It uses a simplified blockchain architecture that is scalable and efficient for lightweight communication systems. The system is composed of three layers: the user layer, the management layer, and the storage layer. In the following, we briefly explain the rules for each layer:

- **The user layer:** The user layer includes the individuals or the organizations who want to access or store their data and services.
- **The management layer:** The management layer includes issuers, verifiers, and consensus nodes. The issuers authenticate the users when they first come and handle their registrations. The verifiers authenticate the users later and manage their keys. The consensus nodes construct the blockchain network and process the new blocks the same way as in Bitcoin processing.
- **The storage layer:** The storage layer includes cloud-based data storage and processing infrastructures to securely store and process the data.

The block structure in the BBDS is simplified by modifying the transaction and block header fields to meet the healthcare records requirements. Furthermore, the interactions in the system are secured by identity-based authentication and encryption techniques which are simple, efficient, provably secure, and lightweight.

BBDS is implemented in a private permissioned blockchain that does not rely on any of the open-source blockchains discussed earlier in Section II-D. The theoretical and the initial implementation of the proposed approach showed a good performance compared to Bitcoin complexity. However, the full system is under development and thus it is not yet popular at the time of this writing.

*3) FairAccess*

FairAccess utilizes the smart contracts to define the access control policies and make authorization decisions [59]. The system uses the blockchain transactions to define authorization tokens. These tokens are used by the sender to authorize the receiver in accessing parts of the sender's data. Functions in FairAccess include: resource registration, grant access, request access and revoke access.

This approach is not implemented in the paper; however, the theoretical analyses showed that the data privacy could be preserved by the provided integrity, authentication, encryption and consensus access control monitoring.

*4) Dynamic Access Control for IoT Using FairAccess*

FairAccess has been utilized to provide a distributed, secured, and adaptive ACL management for the IoT environments [60]. The proposed idea is to let the users register their new resources and define their access policies through the smart contracts associated with these resources. The process of requesting a resource, as depicted in Fig. 7, involves several steps. First, when a request is made for a resource that is held by user A, it is directed to the blockchain network. In turn, the blockchain network allows/denies the access request based on the associated resource's smart contract. The network sends a feedback to the requester granting or denying his access request. Further, the owner can update his/her access policy based on the received feedback from the blockchain network using deep reinforcement learning, an adaptive machine learning mechanism.

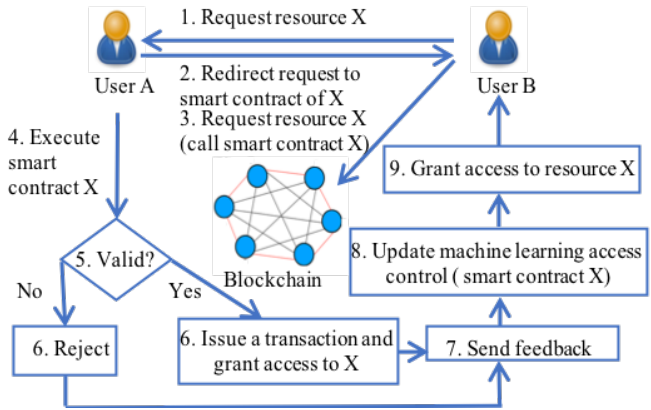

Fig. 7: FairAccess resource request process

This approach is implemented and testified for a specific IoT use-case. The implementation is done on top of Bitcoin and the results show the feasibility of the proposed approach in providing the right access control list management. However, the lack of real-time support, the block complexity and the inflexible implementation are still drawbacks that burden the practicality and the widespread of the proposed approach.

### 5) Decentralized Runtime Access Monitoring System (DRAMS)

In [61], the authors utilize the blockchain technology to verify access control logs for clouds in a federated cloud environment. The key idea is to use the smart contracts in defining the access rights and collecting the access logs from different clouds. The blockchain miners compare the access rights to the access logs. If a violation is detected, an alert is raised to be further handled by the system.

This approach was implemented on top of the Ethereum platform. Results show that the system is resilient to many threats, including compromising the communication channel to modify the access rights, compromising the policy evaluation to allow unauthorized accesses, and compromising the logs to alter or delete them. However, latency, cost and scalability are the challenges that need to be considered for this platform to become practical.

### 6) Data Privacy for IoT Data Storage and Sharing

IoT is witnessing a rapid increase in the number of innovative applications; however, security is still a major concern. Most current ACL mechanisms are delegated to a trusted, centralized controller that maintains and manages the access controls. On the other hand, the blockchain-based data privacy can help resolve the problems with centralization and provide decentralized, resilient, and auditable privacy guarantees. In [62], the authors use the blockchain network to securely store and manage the access permissions. The blockchain transactions are composed of the ownership of the data and the corresponding access permissions. Initially, the owner submits a transaction that includes the data stream identifier. A new transaction is issued when the owner wants to share the data with other users or with the service providers. Further, the owner can revoke his data sharing by submitting a revoke transaction. When a provider or another user wants to retrieve a specific set of data, they send a request to the storage node which queries the blockchain network for the access rights. Data, in both the blockchain database and the storage nodes, is encrypted and highly distributed. This mitigates the threats of malicious storage nodes that grant service access rights without consulting the blockchain.

An initial implementation of this scheme was built on top of the Bitcoin platform. Results showed a reasonable overhead due to the routing loads, the point-to-point communication and the distributed storage. However, the latency and the scalability challenges are not resolved, especially for real-time IoT applications.

### F. Summary and Comparisons

The data privacy is a critical security aspect that guarantees the user's control over their data disclosures and prevents unauthorized access and processing. In this section, we discussed several blockchain-based approaches providing the data privacy. Such approaches define the ACL either by the smart contracts or by special management transactions. Monitoring of the access rules and the violations can be done by the blockchain nodes to fully eliminate centralization. Table VIII compares the different discussed approaches. It should be noted that these approaches handle the data privacy rather than the user privacy. Most blockchain implementations provide a pseudo-anomalous user privacy using the hashes to identify the users rather than their actual names.

TABLE VIII
BLOCKCHAIN-BASED DATA PRIVACY APPROACHES COMPARISON

| Approach | Blockchain platform | Modifies implementation | Smart contract/ transactions | Scalable solution |
|---|---|---|---|---|
| Zyskind | Ethereum (Enigma platform [59]) | Yes | Transactions | No |
| BBDS | Bitcoin | Yes | Transaction | Yes |
| FairAccess | Not implemented (Ethereum is planned) | Yes (planned) | Smart contract | No |
| FairAccess for IoT | Not implemented | Yes (planned) | Smart Contract | No |
| DRAMS | Ethereum | No | Smart contract | No |
| [62] | Yes (Bitcoin) | Yes | Transactions | Yes |

## V. Provenance Services

Data or resource provenance is another security service that deals with the tractability and the auditability of the resources. In this section, we discuss the traditional techniques provide the data provenance and highlight their problems. Following that, we discuss how the blockchains can help in providing a provenance architecture, highlight some of the proposed approaches that utilize the blockchain technology to provide the data provenance services and give a brief comparison among the different discussed blockchain-based provenance approaches.

### A. Data and Resource Provenance

Data provenance refers to the metadata that tracks and reports the originality of the data and the operations associated with them. The metadata includes records of the inputs, the entities, the systems, and the processes that accessed or manipulated the data of interest. An example would be the tracking of the data ownership and the accessing of some information in a cloud environment. When dealing with clouds, the data are massively scaled, and the resources are shared by many different entities. It is important to track the origin of the data and the operations happened on them, including the reading, the processing and the writing of the data or the resources. This is not only applied to the data, but also to any type of resources such as the network devices, the workflows, the web services, and the processes.

Providing provenance guarantees resource tractability, forensic capabilities, and auditability. In other words, it helps the network administrators in detecting any access violation or any malicious operation. However, this service comes with two issues, complexity and privacy violation. Keeping track of the resources is challenging and complex, especially for the distributed applications. Data or resources can be replicated in different areas to provide availability and they might follow different paths to provide resilience guarantees. Furthermore, the amount of data is massively increasing, which makes tracking complex and inefficient. Further, such tracking may violate privacy if the information about the data ownership and the data originality is exposed. Due to this reason, guaranteeing the data provenance without violating the privacy is a challenge to be resolved in the current applications.

### B. Importance of Provenance

In the age of social networking, cloud computing, IoT, and other distributed applications, data is an acute resource that is open and vulnerable to intrusions. The owners need to know not only the data originality, but also the manipulations and the accesses to the data along its lifecycle. For example, in IoT applications, the sensor data has to be tracked so that they get to the consumers without any unauthorized modification. Further, the consumers need to know how accurate the information is and what time it was sent. This can be achieved only by proper data provenance techniques. The same provenance requirements are applied to the healthcare data, the financial data, the governmental resource, or even scientific applications. Such applications are worldwide, generating massive amounts of data that need to be tracked. Hence, the provenance guarantees are crucial for these applications.

### C. The Traditional Techniques

State-of-the-art techniques in providing the data provenance in the cloud environments are based on logging and auditing techniques. Most of these approaches are used at the centralized authority that manages the system resources. Examples of the data provenance approaches include PASS [63], S2Logger [64], and SPROVE [65]. PASS was one of the first approaches to provide the data provenance service by collecting and maintaining information about the operations done at the system level [63]. S2Logger is a tracking tool that provides an end-to-end resource monitoring in a cloud environment at the file level [64]. SPROVE is a technique that provides confidentiality and integrity of the data provenance through encryption and signature techniques [65]. In addition, in [66], the authors propose a secure data provenance technique that utilizes encryption techniques to enhance the privacy of the data provenance.

### D. Problems with the Traditional Techniques

The techniques discussed in the previous subsection have several challenges, including ineffectiveness, complexity, lack of privacy and centralized controllers. The cloud hardware and software are distributed by nature and have several layers of interoperability, which makes the logging techniques inefficient. Tracking resources can be complex in nature as the cloud resources may move to provide load balancing and to ensure resilience. Further, employing security techniques like encryption and digital signature can add an additional level of complexity to the system. However, not having encryption and signature may break the data privacy if the data's origin and data ownership are exposed to a third party. Finally, to store the logging information or to monitor the data in a system, a centralized controller is needed, which requires a trusted third party that is complex, expensive and a single point of failure.

### E. Blockchain-Based Data Provenance

The blockchain technology can be viewed as a shared immutable ledger that record the events in the system. Thus, it is a potential approach to provide the data provenance service by recording the evidence of the data originality and the operations in the blockchain transactions. However, the integrity and the confidentiality of the blocks should be granted by any blockchain-based data provenance. The problems with the traditional approaches and how the blockchains can resolve them are presented in Table IX.

In the following, we discuss several approaches that utilize the blockchains to provide the data provenance service. Further, we discuss some other supply chain provenance approaches and how they can be applied to provide the data provenance service.

TABLE IX
TRADITIONAL DATA PROVENANCE PROBLEMS AND BLOCKCHAIN-BASED SOLUTIONS

| The traditional approach | The problem | Blockchain solution |
|---|---|---|
| Any centralized data provenance technique (PASS, S2Logger, ...) | Ineffectiveness | The distributed nature and the event recording properties of the blockchains can resolve the problem |
| | Complexity | It is delegated to the blockchain network and distributed among the nodes |
| | Data privacy | The sophisticated encryption techniques in the blockchains can help. However, the verification, in this case, would be difficult |
| | Centralized controller | No centralized controller is involved |

### 1) ProvChain

ProvChain is a blockchain-based data provenance system that offloads encrypted provenance records to the blockchain database in the form of transactions [67]. It utilizes the blockchain database as a distributed database that provides the integrity and non-reputability guarantees. The validation of the data provenance is done off-chain by a centralized provenance auditor (PA). The system consists of five components: users, cloud service providers, blockchain network, provenance database, and the PA. The users are the resource owners or the data accessors. The providers offer storage services and are responsible for the users' registration. The blockchain network consists of nodes that participate in the system and keep the data provenance records in the blocks. The provenance database records all the provenance data on the blockchain network and locally at the cloud level. Finally, the auditor retrieves the provenance data from the blockchain database and validates the blockchain receipt. The interactions among the various system components are depicted in Fig. 8.

ProvChain has been implemented and evaluated in [67]. The results demonstrated the feasibility of the proposed approach and the ability to supply all the security features provides by the blockchains. ProvChain utilizes a blockchain database to store the data provenance records. However, the verification of the operations and the records is done locally outside the blockchain network, at a centralized PA. In other words, ProvChain does not utilize the consensus property of the blockchain technology.

### 2) DataProv

DataProv is another platform that uses the blockchain technology and the smart contracts to provide data provenance services for the sensitive cloud information [68]. It is built on top of the Ethereum platform and uses an off-chain JavaScript module to interact with the users. There are two types of smart contracts: Document_Track and Vote. A Document_Track contract is an Ethereum smart contract and is initiated for each document in the system. The documents can be shared data or any type of assets shared by the system. The *Document_Track* contract includes functions such as add a document, grant a user access, revoke an access, and track the changes in the document. The *Vote* contract records the miners' votes and includes functions such as initiate a vote, record a vote and terminate a process. An overview of the system architecture is illustrated in Fig. 9. Initially, the user submits any changes in the data to the blockchain network and submits a vote contract. The blockchain network asks the miners, or the voters, of the system to verify the changes and to report back. The voting is done based on the majority votes from the miners involved and the votes are recorded for each change. At the end of the voting phase, the changes to the document are either accepted or rejected. If the changes are accepted by the voters, then the Document_Track contract is updated, and the changes are submitted to the cloud provider. Otherwise, the changes are rejected after some time and the cloud provider will be notified to take further actions.

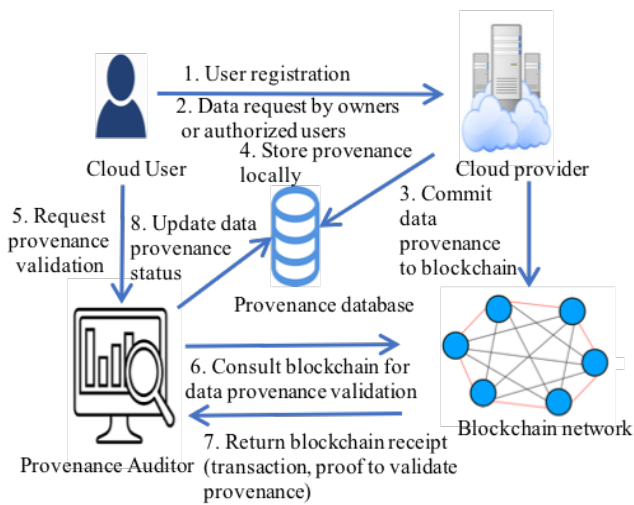

Fig. 8: ProvChain system interactions

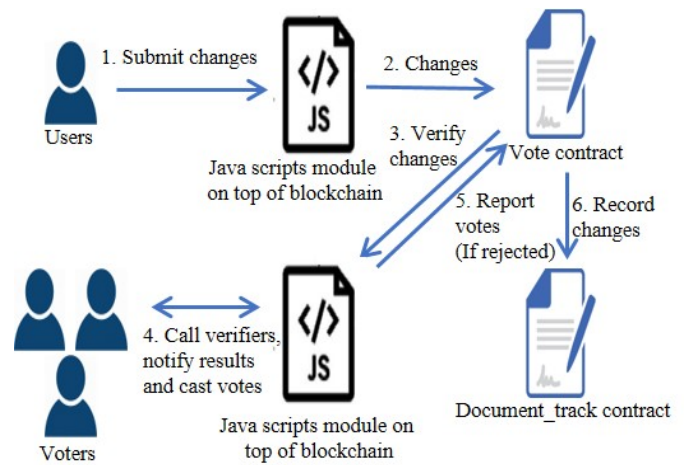

Fig. 9: DataProv architecture

DataProv has been implemented on top of the Ethereum blockchain platform. Drug trials and wheat production are the specific use-cases used to evaluate the system. The evaluation is done on real-life scenarios and shows the feasibility of the proposed approach in providing provenance guarantees with low cost and moderate overhead.

*3) A Blockchain-Based Approach for Data Accountability and Provenance Tracking*

In [69], Neisse et al. propose a blockchain-based approach for the data accountability and the provenance tracking. Similar to DataProv, this approach uses the smart contracts in the public blockchains to define the access rules for the data. This system is composed of three main actors: data subjects, data controllers, and data processors. Data subjects are the owners of the data and they authorize the data controllers to access the data. Meanwhile, data processors are the organizations that are authorized to process the data on behalf of the controllers. This approach defines three types of smart contracts that can be summarized as follows:

- **The data subject contract for specific controllers:** The owner of the data creates a contract for each specific controller that has all or portions of the data. This contract tracks the data shared with the controller, the access rights and the operations performed by the controller on the data. Hence, this contract provides the provenance service of the data shared with that specific controller.
- **The data subject contract for specific data:** The owner defines the rules and the access rights of a set of data that is accessed by any controller in a smart contract. This contract tracks a specific data shared with any controller and the logs associated with operations performed on that data. In other words, it provides provenance for a specific data or a set of data that can be shared with any controller.
- **The data controller contract for data subjects:** The controller creates a contract for any owner that wants to share the same data with it. This contract defines how the data received from the data subjects are treated by the controller. It is used by the subjects, or the owners of the data, to help decide data sharing rules with that controller. Further, the contract logs the creation and ownership of the data, which is also a part of the data provenance guarantees.

These contracts provide both privacy and provenance for the data shared with the clouds, the grids or even the services shared with others. Thus, the approach can be used for any security service; For example, sharing the patient's personal information in a smart healthcare system.

Initial implementation showed the feasibility and practicality of the proposed approach in providing provenance and privacy. However, the implementation also showed scalability and performance limitations due to the blockchain's complexity.

*4) Provenance (An Enterprise Project)*

Provenance is an enterprise project that offers a blockchain-based tracking of physical items or resources in a supply chain [70]. It associates every physical product with a digital identity to provide the traceability and the transparency for that product. The idea is to have the producers register their products to the blockchain system with certifications or tags. Then, the blockchain network tracks the changes made to that product. An event is recorded to the blockchain each time a change is made, or a product reaches a certain stage in its lifecycle. In this way, the products can be traced when shipped from the producers to the consumers.

Use-cases showed the feasibility of the Provenance's approach to track commercial products such as foods or goods. Even though Provenance's technique strictly applies to the physical products, it can be extended to audit and risk management of the data in enterprises. For example, it is possible to track the data in the cloud when they move from the data center to the user or vice versa. In this case, Provenance can provide a solution by letting the user register the data in the blockchain, where the tracking, the recording and the validation of the data can happen.

*5) IBM Supply Chain*

Similar to Provenance, IBM provides their own blockchain-based supply chain tracking using the Hyperledger blockchain platform. The concept is similar to other blockchain-based data provenance, whereas any change or any operation done on the data is logged to the blockchain database in transactions. IBM use-cases are also dedicated to tracking physical products; however, they can be extended to digital applications easily [71].

*6) Other Blockchain-Based Supply Chain*

Even though IBM and Provenance are the most popular blockchain-based supply chains, there exist some other approaches that do the same. These approaches mostly care about the physical products and other goods provenance, but their approach can be applied to the data and the digital resources, as in Provenance. BlockVerify is a startup that tracks counterfeit products to detect frauds. Their main use-cases include the pharmaceutical industry, luxury products, diamonds, and electronics [72]. Ambrosus is a startup company that provides supply chain provenance, mainly for the food products and medicines [73]. Moreover, EverLedger is another startup company that provides provenance guarantees for digital and physical products using a similar concept [72].

*F. Summary and Comparisons*

With the amount of data that are being processed in the current applications, it is critical to know and understand the data originality, validity, and timing. Data provenance is one way to provide that by tracking the data ownerships and recording the changes. Provenance is critical for current applications' auditing and the detection of security violations. The traditional techniques are mostly inefficient, complex, centralized, and have no specific protection for sensitive information. In this section, we discussed blockchain-based data provenance and described several approaches to achieve the data provenance with the blockchains. Further, the blockchain-based supply chain provenance can be modified to provide data provenance in multi-cloud environments. Table X compares these approaches.

The complexity of the communication in DataProv and the centralized PA in ProvChain show that the blockchain-based data provenance needs more research effort and can be further enhanced by proper utilization of the smart contracts.

TABLE X
BLOCKCHAIN-BASED DATA PROVENANCE APPROACHES COMPARISON

| Approach | Blockchain platform | Modifies implementation? | Data or physical supply chain |
|---|---|---|---|
| ProvChain | Bitcoin | No | Data – single cloud |
| DataProv | Ethereum | No | Data - multiple clouds |
| Neisse [69] | Their own | Yes | Data- any type of data sharing |
| Provenance | Ethereum (mostly but not stated) | No | Physical supply chain mostly-discussed Auditing and management of data for enterprises |
| IBM-Supply Chain | Hyperledger | No | Physical supply chain but can be extended to multiple clouds |
| BlockVerify | Their own | Yes | Physical supply chain |
| Ambrosus | Ethereum | No | Physical supply chain |
| EverLedger | Their own | Yes | Both physical and digital supply chain |

## VI. INTEGRITY ASSURANCE SERVICE

Integrity assures that the data has not been modified or altered when at rest or in motion. In this section, we discuss the integrity service, its importance for the current applications, its traditional techniques, and the challenges associated with them. Then we highlight how the blockchain technology can help in the integrity verification and discuss some of the proposed blockchain-based integrity assurance platforms.

### A. Integrity Assurance

Integrity assurance deals with the correctness and the validity of the data stored, accessed, or generated by the network. It assures that the information has not been changed or corrupted by unauthorized users. This should be applied to the information in motion or at rest. In other words, when the data is stored in the cloud, generated by a sensor, or is transmitted to a client, it should not be altered by an unauthorized user.

Providing an end-to-end integrity assurance maintains consistency, reliability, accuracy, and trustworthiness of the information over its entire lifecycle. The integrity is one of the basic components of the CIA (confidentiality, integrity and availability) triad for information security [74]. Further, it is a required service by any interconnection system, as was discussed in Section I-B [15]. Therefore, guaranteeing the integrity service has been investigated for several decades. In most cases, integrity is ensured by proper signatures and public key cryptographic techniques [28]. However, how to define and trust a third-party authority to verify the integrity is a challenge, especially in the current distributed networks.

### B. Integrity's Importance for The Current Applications

Current applications deal with massive amounts of remote communications that involve many actors, multiple intermediate devices, and several domains. This makes the data, the users, and the information inextricably linked to the cyberspace and vulnerable to many attacks. Attacks include data thefts and alterations, which might threaten human lives. For example, in the smart healthcare system, altering the patients' information can result in serious consequences, if the sensor data is altered by intruders. The same threat can target the IoT platforms, the smart home environments, or the intelligent transportation systems. Thus, it is critical to provide integrity for the information shared by the network. Further, it is important to know the source of the alteration and react to the changes as quickly as possible.

### C. Traditional Integrity Assurance Techniques

Data integrity is commonly assured using cryptographic tools and data replications. Cryptographic tools such as the public key cryptography or the keyless signature infrastructure (KSI) are used to sign the data or the resources so that an unauthorized person cannot change them. Any change in the data will be detected by the signature validation techniques. The process is similar to entity authentication which was explained in Section III-A. Attacks on such techniques require the attacker to know the secret key in order to sign the data. Finding the secret key is challenging, but once realized, the attack becomes practically unpredictable. Replications can also protect against the data integrity violations. In this technique, the attacker needs to modify all the replicas, which can be distributed over several nodes in a randomized fashion. Thus, combining both the replication techniques and the cryptographic tools can provide the system with a strong a data integrity assurance. These are commonly used nowadays in the cloud environments [75].

### D. Problems with the Traditional Techniques

One of the most critical problems with the previously discussed techniques is not in verifying the integrity but is in tracking the intruder who tampered the data. It is not feasible in practice to find the intruder as the tampering could be done at the storage phase, at the processing phase, or at the communication phase. However, knowing the intruders can help in detecting malicious behaviors, changing the access control mechanisms and, in some cases, penalizing these intruders. In addition, the validation of the data is currently done by a trusted party, which imposes a security risk of trust and a single point of failure. Further, integrity is normally an additional security service which adds work, resources, and complexity to the system.

### E. Blockchain-Based Integrity Assurance

Blockchain-based architectures are potential approaches to solve the problems discussed in the previous subsection. The problems with the traditional approaches and how the blockchains can resolve them are illustrated in Table XI.

TABLE XI
PROBLEMS WITH THE TRADITIONAL INTEGRITY
TECHNIQUES AND BLOCKCHAIN-BASED SOLUTIONS

| The traditional approach | The problem | The blockchain solution |
|---|---|---|
| Cryptography and Replication Techniques | No Tracking | The blockchains can be used to save data changes, thus, they provide the evidence to the latest changes (as discussed in data provenance section) |
| | Integrity is an additional service | The blockchains include the transaction integrity check by design, thus, the integrity is not an added service. |
| | Outsider tampering | The blockchain database cannot be tampered |

The blockchains have embedded integrity checks as transactions are signed by the sender and verified by the miners. The data cannot be tampered if it is committed to the blockchain database as discussed before. Thus, the use of the blockchain transactions to submit the data or any asset guarantees the integrity service. Moreover, the blockchain technology can be used to provide evidence for when the data has changed as discussed previously in Section V on the data provenance. In the following, we discuss several integrity techniques using the blockchain technology.

*1) Blockchain-Based Data Integrity Service Framework for IoT Data*

In [76], the authors propose a blockchain-based data integrity framework that uses the smart contracts to achieve its objectives. This framework is dedicated to IoT applications that require a producer-consumer architecture. In this architecture, the owner shares the data with other consumers for specific purposes. The data is generally shared through the use of the cloud storage services, where the owner posts the data to the cloud and the consumers access the data from there. As discussed in the previous subsection, storing the data in the blockchain database provides the integrity service. However, the blockchain database are limited in memory and cannot handle the massive amounts of data. Thus, storing all the cloud data becomes impractical.

The idea of this framework is to store encrypted hash values of the data on the blockchain database and these hash values are then used to check the integrity. The owner generates the hash value of the data, encrypts the hash value and sends it to the blockchain network as a smart contract or a transaction. Further, the owner posts the data to the cloud and allows other users to access it. The procedure of the integrity assurance is as follows. First, the owner or the consumer requests the cloud storage to provide the data stored in the cloud. The consumer calculates the hash value of the retrieved data. Alternatively, the owner could ask for the hash value directly from the cloud if the cloud is capable of doing hash calculations. Then, the owner or the consumer consults the network for the hash value of the same data. If the hash values from the cloud and the blockchain response match, the data is valid, otherwise, it is not valid.

This approach has been implemented on a private blockchain. The initial results showed that this technique can support the integrity verification efficiently in a small-scale network. However, as it can be seen, the proposed framework uses the blockchain technology as a distributed database and it does not make use of the most appealing blockchain characteristics. Furthermore, the consumer could be an IoT device which generally lacks the required computational power and might be unable to perform all the required computations. A better approach would delegate the construction and the validation of the hash values to the blockchain network, which is consulted for the validity of the data.

*2) Storj*

Storj is a blockchain-based peer to peer data storage system that utilizes the blockchain database to store hash values of the data and verify the integrity. The blockchain technology is immutable and provides integrity checks by design. Thus, any storage system that utilizes the blockchain transactions to store the data can provide the integrity service. The data can be stored in the blockchain transactions or off-chain by storing some metadata in the transactions and the data itself in the off-chain storages. In Storj, the data is stored off-chain while the metadata referring to the original data is stored in the transactions. The metadata has the location of the data and the hash of the data. Whenever the user wants to access the data, it inquires the blockchain network. The network validates the data stored off-chain and returns back the metadata needed to retrieve the original data. In this way, the integrity is provided efficiently; however, the requirement of tracking the intruders in case the data is changed is still not provided [9].

Storj is an open-source implementation that allows secured and integrity-guaranteed data storage in distributed applications. It has been verified and tested by many real-case scenarios. To the best of our knowledge, Storj is the first blockchain-based cloud storage that is currently used by some enterprises.

*3) Ericsson Blockchain-Based Integrity Assurance*

Ericsson partnered with Guardtime to provide integrity services that allow the application developers to assure the integrity of their users' data and assets. They utilize *Keyless Signature Infrastructure* (KSI) to generate signatures for the resources [77]. KSI is a signature technique proposed in 2006 exploiting hash trees and timestamps to construct a signature for multiple documents. However, the original proposal relied on a central authority to construct the tree and give the signatures. Guardtime provides a blockchain-based KSI approach which is scalable, decentralized, efficient and provably secure. The Guardtime solution can be used for the authentication as it was discussed in Section III or for the integrity assurance. Ericsson utilizes Guardtime to provide the integrity rather than the authentication. The basic objective is to verify that a collection of the data generated by an application has not been modified. The functions provided by the service include generating a non-invertible signature for the user's data, extending the signature or publicizing it, and verifying the integrity of the users' data [78].

The Ericsson service involves two simple steps: signing the data and verifying the signature. The signature of the data is recognized by submitting them to the blockchain, where the signature is simply sent back to the user. The signature is stored in the blockchain transactions as well as at the user's system. To verify the data, the stored signature is submitted to the blockchain network for verification purposes. The blockchain nodes will validate the signature and return the expected hash value if the signature is valid. The user compares its hash value with the submitted one to determine whether the data was modified [79].

This service is provided as an open-source software development kit used by the application developers. It is currently used in Ericsson cloud solution to provide immutable evidence for the data stored in their clouds. This service is expected to get more popular and adopted by other cloud providers to provide a blockchain-based integrity assurance.

### F. Summary

Data integrity assures that the data stored or transferred has not been tampered. It is normally done by the cryptographic signature combined with the verification techniques. In this section, we discussed the data integrity assurance using the blockchain technology and highlighted some blockchain-based integrity assurance approaches. The blockchain technology by itself can provide integrity assurance through non-repudiation guarantees. Table XII compares the different approaches from several perspectives. The Ericsson service verifies that a collection of data has not been altered by storing their signature on a blockchain.

TABLE XII
BLOCKCHAIN-BASED INTEGRITY ASSURANCE APPROACHES COMPARISON

| Approach | Blockchain platform | Used cryptographic approach | Can verify multiple chunks |
|---|---|---|---|
| Liu [76] | Private blockchain | PKI | No |
| Storj | Florincoin [80] | PKI | No |
| Ericson | Guardtime | KSI | Yes |

It should be noted that none of the proposed approaches guarantees tracking the intruders if the data has been changed. Combining the proposed approaches with the data provenance approaches discussed earlier can provide tracking of the data; thus, it can detect who was the last changing the data.

## VII. BLOCKCHAIN CHALLENGES

Despite the potential benefits of the blockchain technology, it still has some challenges that limit its practicality for the security applications discussed in the previous sections. In this section, we highlight some of these challenges and relate them to the security applications studied in this paper.

### A. Privacy and Anonymity

One of the blockchain's main properties and advantages is providing pseudo-user anonymity. This is critical for security as the public blockchains are open and the user information would be exposed to attackers. However, for most of the discussed approaches, the transactions relate the user identity to their public key, the ACL, or the provenance data. For example, the blockchain-based ACL mechanisms relate the ACL to the users directly; therefore, the users are no longer anonymous. The same issue is applied to the blockchain-based key management and blockchain-based provenance. That is, the privacy and the anonymity features of the blockchains are flawn. Bitcoin resolves the anonymity problem by using the user's public key as the user identification. However, this provides pseudo-anonymity and further research is needed to provide fully anonymized approaches that meet the security application requirements.

### B. Computations and Mining Nodes

In most of the current applications, the nodes are simple and do not have high computational capabilities. That is, the blockchain clients need to be simple in order to satisfy the low computation capability requirements. On the other hand, the security services, in general, require significant computations in encryption, decryption, and signature. Moreover, as discussed in Section II, the blockchain technology needs to have mining nodes with high computational power. For most of the proposed techniques, the mining challenge would be resolved by allowing the application nodes to be the blockchain clients and by introducing dedicated mining nodes that are added just to perform mining. However, the high computational power required for these nodes adds to the cost of the system. A better approach would include reducing the computational needs for the mining and relating the mining powers to the node trustworthiness or its reputation in the system. Further, simpler cryptographic schemes can be developed to reduce the computational needs for signing and encrypting the data.

### C. Communication Overhead

Current applications are highly dynamic; therefore, they require frequent changes in the access lists and the provenance data. This forces the nodes to send frequent transactions to update the ACL or modify the provenance information. On the other hand, the blockchain technology is a peer-to-peer network, which indicates that a significant overhead will be added in terms of the network traffic and the system processing capabilities. The transactions and the blocks need to be broadcast as opposed to unicast in the traditional techniques. Thus, the overhead added to the network is significant and a considerable challenge. The storage and the processing overhead bring additional challenges in adopting the blockchains for security applications.

### D. Scalability

The blockchain technology is believed to scale better than the traditional centralized techniques. However, as reported in [81], the technology performs poorly as the number of users and networking nodes increases [81]. This is a major challenge, especially with network security applications, where thousands of users need to be served and the network scales up fast. Furthermore, the dynamicity of the system adds to the scaling problem as the nodes need to frequently send update transactions. The Ethereum platform and the Hyperledger platform have their own promises for scalability. However, the performance tests done in [24] show that both platforms still suffer from some aspects of scalability issues.

*E. Time Consumption*

Providing security services requires fast processing capabilities, especially in the current networks, where milliseconds can cost billions of dollars. Further, mining and achieving consensus are still time-consuming in the blockchains. The proposed approaches resolve the problem by making decisions from the local blockchain logs without requiring distributed consensus. For example, in the blockchain-based ACL mechanisms, the access decisions are made based on the local copies of the blockchain database. However, this defeats the technology decentralized architecture and its consensus as the nodes need to trust the local blockchain database and make centralized decisions. Many promises have been made to resolve Bitcoin's time issues in Ethereum and Hyperledger platforms. However, the time required for mining is still two or three seconds as compared to the milliseconds requirement. Furthermore, building encryptions and security techniques over the blockchains exacerbates the problem of time complexity since such techniques are complex and time-consuming. Thus, faster mining and processing techniques are needed to be able to employ the blockchains for real-time applications.

*F. Summary*

The popularity of the blockchain technology in several nonfinancial applications raised multiple challenges that we discussed in this section. The discussed challenges are related to providing security services and meeting the requirements of the current applications. These challenges include privacy and anonymity, computations and mining nodes, communication overhead, scalability, and time consumption. Privacy and scalability are the most difficult challenges, since they are related to the blockchain-based security applications. A balance between the technology potentials and the its challenges should be considered for efficient designs and solutions. Table XIII summarizes the blockchain-based security application challenges. Till now, the blockchain technology does not seem to be a potential candidate for real-time and delay-sensitive applications. Thus, the future research should tackle these challenges for a practical and widespread use of the blockchain-based security applications.

I. CONCLUSIONS

In this paper, we presented a comprehensive survey on the utilization of the blockchain technology in providing distributed security services. These services include entity authentication, confidentiality, privacy, provenance, and integrity assurances. The entity authentication and the confidentiality can be achieved by the public key cryptography using encryption and the signature schemes. Thus, we discussed different blockchain-based key management for public key cryptography. Further, privacy, provenance, and integrity assurance services were studied each in separate sections.

TABLE XIII
BLOCKCHAIN-BASED SECURITY APPLICATION CHALLENGES AND PROPOSED SOLUTIONS

| The challenge | The problem | Effect on security for current applications |
|---|---|---|
| Privacy and anonymity | Most of the proposed approaches relate the user identity to the user information. For example, a user and its access control list. | Breaks user anonymity since the user information is exposed and it is possible to relate the user identity to their information. |
| Computations and mining nodes | Signature, encryption and mining require high computational power which is not feasible with the resource limited application. | Most of the current application are resource limited, thus, the proposed security management is not feasible without adding extra nodes for mining and simplifying the signature schemes. |
| Communication overhead | The blockchain network is a peer-to-peer network which imposes high overhead in term network traffic and processing. | As current applications are highly dynamic, the ACL and the data provenance frequently change. This indicates that the nodes will be sending lots of update transactions which will magnify the overhead even more. |
| Scalability | Bitcoin, Hyperledger and Ethereum have scalability issues as was indicated by [81]. | Most of the proposed approaches were applied to limited scale networks. Testing them to real applications could be a challenge, especially for the blockchain-based ACL and the blockchain-based provenance. |
| Time consumption | The time required to do mining and reach consensus could be high especially for real-time applications. | The security applications need fast processing capabilities especially for the current resource constraint applications. |

We summarized on the properties that make the blockchain technology a potential candidate for several distributed applications. Then, we defined each service, discussed its rules in the current networking applications, highlighted the traditional approaches achieving the required service along with their challenges. Finally, we explained how the blockchains can help resolve these problems; explored different blockchain-based approaches and presented a comparison of such approaches. At the end, we studied the challenges that are currently restricting the blockchain's practicality for security applications. The blockchain technology seems to have a great potential in many applications; however, its practicality in security applications is still questionable due to several challenges. Future research directions include resolving these challenges and testing the different blockchain approaches in large scale and real-time environments.

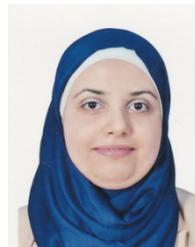

Tara Salman is an IEEE student member. She received her BS and MS from Qatar University Doha, Qatar at 2012 and 2015, respectively. Her BS was in computer engineering while her MS was in computing (networking minor). She is currently pursuing a PhD at Computer Science & Engineering at Washington University in St Louis, Missouri, USA.

From 2012 -2015, she worked as a research assistant at Qatar University on a NPRP (National Priorities Research Program) funded project targeting physical layer security. She is working as a Graduate Research Assistant at Washington University in St. Louis since 2015. Her research interest spans network security, distributed systems, Internet of Things and financial technologies. She is an author of 1 book chapter, 6 research articles and has been a presenter at many international conferences.

Salman is a recipient of the Cisco Certified Network Associate (CCNA) certification in 2012 and had completed all CCNA academy levels at Cisco-Academy (Qatar University branch).

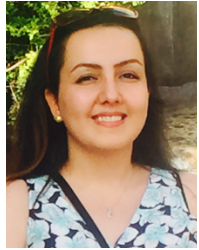

Maede Zolanvari is an IEEE student member. She received her B.S. degree in Electrical and Computer Engineering from Shiraz University, Iran, in 2012 and her M.S. degree in Electrical and Computer Engineering-Communication from the same university, in 2015. She is currently a Ph.D. candidate in computer science and engineering at Washington University, St. Louis, MO, USA.

She was a member of Gifted Talent Office at NODET (National Organization for Development of Exceptional Talents) during her education at Shiraz University. Since 2015, she has been working as a Graduate Research Assistant at Washington University in St. Louis. Her research interests include secure computer networks, wireless communications, and Internet of Things.

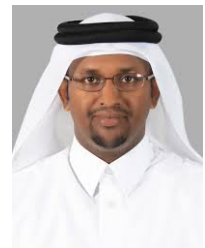

Dr. Aiman Erbad is an Assistant Professor at the Computer Science and Engineering (CSE) Department at Qatar University. Dr. Erbad obtained a PhD in Computer Science from the University of British Columbia (Canada) in 2012, a Master of Computer Science in Embedded Systems and Robotics from the University of Essex (UK), and a Bachelor of Science in Computer Engineering from the University of Washington (USA). Since September 2016, Dr. Erbad has been the Director of Research Support, responsible for all research grants and contracts. Prior to that Dr. Erbad was the Coordinator of the Computer Engineering program and the Chair of the Curriculum and Quality Assurance committee leading to ABET accreditation and curriculum enhancement efforts at the CSE department.

Dr. Erbad received the Platinum award from H.H. The Emir Sheikh Tamim bin Hamad Al Thani at the Education Excellence Day 2013 (PhD category) and graduated from Qatar Leadership Center, which trains rising leaders in different sectors. Dr. Erbad's research interests span cloud computing, multimedia systems, networking, and security. Dr. Erbad's research received funding from the Qatar National Research Fund and his research is published in reputed international conferences and journals. Dr. Erbad is a member of various University committees (Policy, Ranking, Institutional Effective, Intellectual Property, Appeal and Re-instatement) and the Chair of the University Research Support Committee. He serves as an Editor in the European Alliance for Innovation (EAI) Endorsed Transactions on Collaborative Computing, and as a technical program committee member in various IEEE and ACM international conferences. Dr. Erbad acts as an expert in information technology strategy and research techniques for various national entities.

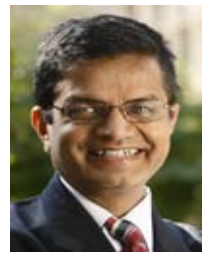

Raj Jain is a Fellow of IEEE, a Fellow of ACM, and a Fellow of AAAS. He received BS degree in Electrical Engineering from APS University in Rewa, India in 1972 and MS in Computer Science & Controls from 32 IISc, Bangalore, India in 1974 and the Ph.D. degree in Applied Math/Computer Science from Harvard University in 1978.

He is currently the Barbara J. and Jerome R. Cox, Jr., Professor of Computer Science and Engineering at Washington University in St. Louis. Previously, he was one of the Co-founders of Nayna Networks, Inc - a next generation telecommunications systems company in San Jose, CA. He was a Senior Consulting Engineer at Digital Equipment Corporation in Littleton, Mass and then a professor of Computer and Information Sciences at Ohio State University in Columbus, Ohio.

Dr. Jain is the winner of the 2017 ACM SIGCOMM Life-Time Achievement Award, the 2015 A.A. Michelson Award, the 2006 ACM SIGCOMM Test of Time award, the CDAC-ACCS Foundation Award 2009, the IISc Distinguished Alumnus Award 2014, the WiMAX Forum Individual Contribution Award 2008, and ranks among the Most Cited Authors in Computer Science.

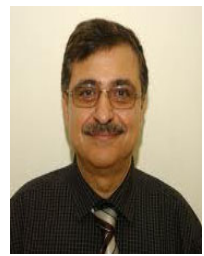

Mohammed Samaka is an associate professor of Computer Science in the Department of Computer Science and Engineering (CSE), College of Engineering at Qatar University. He obtained his PhD and Master from Loughborough University in England and a Post Graduate Diploma in Computing from Dundee University in Scotland. He obtained his Bachelor from Baghdad University.